*Original Article*

# Safe Routing Approach by Identifying and Subsequently Eliminating the Attacks in MANET

S.M. Udhaya Sankar[1], D. Dhinakaran[2], C. Cathrin Deboral[3], M. Ramakrishnan[4]

[1,2,3]Department of Information Technology, Velammal Institute of Technology, Chennai, India
[4]Department of Computer Applications, Madurai Kamaraj University, Madurai, India

[1]Corresponding Author : udhaya3@gmail.com

**Abstract** - *Wireless networks that are decentralized and communicate without using existing infrastructure are known as mobile ad-hoc networks. The most common sorts of threats and attacks can affect MANETs. Therefore, it is advised to utilize intrusion detection, which controls the system to detect additional security issues. Monitoring is essential to avoid attacks and provide extra protection against unauthorized access. Although the current solutions have been designed to defeat the attack nodes, they still require additional hardware, have considerable delivery delays, do not offer high throughput or packet delivery ratios, or do not do so without using more energy. The capability of a mobile node to forward packets, which is dependent on the platform's life quality, may be impacted by the absence of the network node power source. We developed the Safe Routing Approach (SRA), which uses behaviour analysis to track and monitor attackers who discard packets during the route discovery process. The attacking node recognition system is made for irregular routing node detection to protect the controller network's usual properties from becoming recognized as an attack node. The suggested method examines the nearby attack nodes and conceals the trusted node in the routing pathway. The path is instantly assigned after the initial discovery of trust nodes based on each node's strength value. It extends the network's life span and reduces packet loss. In terms of Packet Delivery Ratio (PDR), energy consumption, network performance, and detection of attack nodes, the suggested approach is contrasted with AIS, ZIDS, and Improved AODV. The findings demonstrate that the recommended strategy performs superior in terms of PDR, residual energy, and network throughput.*

## 1. Introduction

A collection of network nodes that interact outside of a fixed physical structure make up a mobile ad hoc network (MANET). MANETs include a variety of noteworthy characteristics, including variable topology, quick deployment, and multi-hop wireless transmission [1]. MANET is appropriate for various time-sensitive tasks thanks to all these qualities. In cases when it is challenging to construct basic infrastructure, ad hoc networks offer a viable communication facility. Additionally, with MANETs, mobile nodes communicate without requiring a physical infrastructure and without engaging in administrative tasks. As a result, these systems are flexible, self-organizing, and automatically created, enabling nodes to move at will while interacting [2-4]. These systems can be expanded and merged to create intelligent solutions for addressing industrial demands with cutting-edge technology like cloud computing, the IoT, and machine learning methods. Figure 1 shows MANET's organizational structure. Given the changeable topology of the MANET, implementing a secure routing protocol while assuring the quality of service is challenging. MANET's open communication technology allows any adaptable node to join the network and participate in data transmission. In MANET, this adequate communication infrastructure leads to security lapses [5-8].

As the traditional routing techniques for these networks presuppose collaborative, trusting settings among mobile devices, secure routing in ad-hoc networks was one of the main challenges for researchers. Attackers can readily hack authorized corporate mobile nodes to cause a variety of packet transmission errors and launch DoS assaults. Sequence number assaults are a typical denial-of-service attack against traditional MANET routing protocols, and they can drop data packets during the transmission process after violating protocol requirements during the route-finding phase. As a result, it is challenging to provide secure and trustworthy routing in such a network. According to their architecture and routing method, routing protocols in MANETs can typically be divided into three categories: proactive routing protocols, reactive routing protocols, and hybrid routing protocols. The MANET's most popular reactive routing mechanism is the ad hoc on-demand distance vector routing protocol (AODV) [47]. Numerous researchers investigated AODV's effectiveness by considering several variables and also discovered numerous causes of security breaches. As a result, these shortcomings of this protocol exist:

- There is no system in place to deal with congestion and packet losses.
- The current protocol does not support multipath routing.
- It is susceptible to numerous security attacks.
- It lacks a predetermined process for dealing with node transmission delay.
- It lacks any safeguards to guarantee service quality.
- The node's energy usage is substantial, and there is no notion of power optimization.

By developing a secure attack detection and recognition that takes into account issues like a safe path to endpoints for transmission of data, negligible node energy usage, and the innovation of protection schemes that really can come to terms with egoistic and harmful attacks on the entire network, the security concerns in MANET can be resolved [10-15]. The assault detection mechanism for the MANET has not yet been implemented completely. The Secure attack gesture recognition algorithm has a distinct structure and design compared to wired and wireless networks. Because MANET lacks a system, node assaults on the networks are unaffected. To illustrate the situations for the needs of the safe assault detection method, Fig. 1 shows the MANET topology with the source and destination nodes, connections in a network, and malicious nodes.

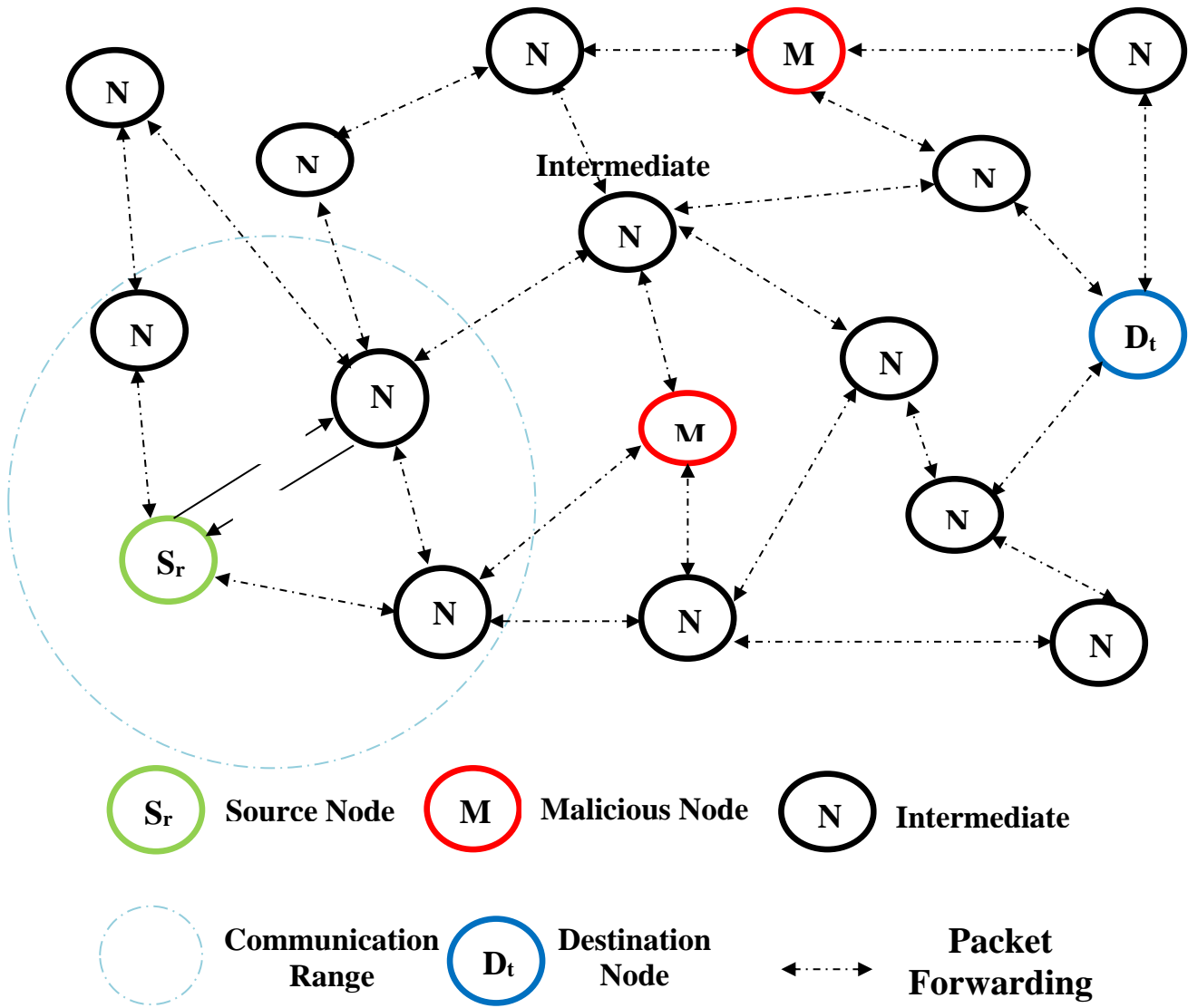

**Fig. 1 The MANET Architecture**

Path computing and node prevention must both be considered by the protocol. The protocols search the network for the most accurate and efficient approach to sink nodes. While placing nodes on the shortest paths, it must also be considered to use the least amount of energy possible [16]. The protocols may allow for multiple ways; if one way breaks for whatever reason, the data will still be sent to the channel's destination via a different path. If dubious nodes have not already been identified in the system, a mechanism must be developed to place them there. To defend against various attacks and keep malicious nodes out of the network's routing path, the Secure assault detection method implementations promptly update node info [27-32].

Ad hoc On-Demand Distance Vector is highly vulnerable to numerous assaults, including wormholes, denial of service, and black hole assaults. It is vital to develop a new strategy to address these security vulnerabilities [33-36]. Numerous scholars have suggested various variations of the AODV protocol to address the abovementioned problems. However, there are currently no AODV protocols that address all the abovementioned problems as a solitary framework. The most recent research findings based on the Mobile Ad-Hoc Network (MANET) security aspect have offered various elements [37]. To ensure a high level of safety in packet forwarding, current research has focused on specific attacks and some approaches that significantly increase communication costs. Based on the findings of the latest analysis, the primary goals were set for creating the proposed methodology with the overarching goal of offering a high level of protection to the content in the MANET [38].

- To develop a reliable and secure navigation system that tackles issues with creating a secure path among MANET nodes.
- To manage and minimize packet losses.
- To provide a system that allows nodes to interact with one another about data securely.
- Finding and ultimately stopping the threats in MANET.
- Improving MANET's level of service quality.

The following parameters for safe routing were taken into consideration when developing the SRA procedure in this paper:

- Secure routing using dependable nodes
- Monitoring Packet Flow.
- Watching over nearest neighbours.
- Discovery of malicious nodes.
- Separate Attack and Reliable Nodes.
- Safe routing prolongs network life, boosts throughput, uses less energy, and reduces packet loss.

## 2. Related Work

Due to its transparency, the changing topology of the network, and the absence of centralized control and monitoring, wireless communication are more vulnerable to attacks than wired ones. In wireless connections, security concerns are becoming ever more crucial. Some considerations should be made when creating an attack node detection mechanism for MANETs. The assault node detection techniques for MANETs will function differently from their wired equivalents. Some issues must be resolved while creating attack node detection techniques for MANETs.

The characteristics of privacy have been covered in several wireless security studies. Numerous studies on protected transit in Mobile Ad-Hoc Networks have been done throughout the last few generations. Since energy constraints and unforeseen events in network activity frequently and unpredictably lead to topological changes, the task of locating and conditioning will lead to WSNs is challenging. The biggest obstacle is figuring out where the agents are while the nodes move. Similarly, the nodes housing the intrusion-detecting agents need more computing power, connectivity, and battery life.

For the scattered context in mobile ad hoc networks, Filipek and Hudec [15] propose a secure structure based on a PKI, trust model, firewall (MANETs), and intrusion prevention system (IPS). Nodes' L0 through L3 of privileges are defined by their trust level, which can be lowered or removed in response to harmful activity. Only the ability to obtain the credential is granted to L0-level nodes. The channel's end-to-end connectivity is permitted with L1-level nodes. Nodes at the L2 layer can participate in distributed, IPS, and forwarding. A stand-alone attribution authority (AA) called an L3-level router can accredit other networks and develop its environment. Using the AA key pair, each node can use the certificate's AA signature to confirm its authenticity. The suggested architecture uses a gateway as a privacy overlay, PKI to impose secrecy and data communication standards, and IPS to govern nodes and ensure compliance with security protocols (PKI and firewall rules). However, DHT-based connectivity is not taken into account by the researchers.

The RAD protocol was created by Hamamreh et al. [16] and utilizes the reinforcing technique, Diffie-Hellman, and MD5 for communication security. This method examines node behaviour in the MANET using a learning algorithm. This strategy locates and removes the network's harmful nodes. The Diffie-Hellman algorithm is employed to share private keys amongst some nodes, while MD5 is utilized to identify among nodes. This RAD method does not necessitate a third party to disseminate secret keys. Additionally, this protocol prevents choosing a path containing malicious activity.

Zarei and Fotohi [17] presented a thorough architecture for the technology of the IoT's centered on the SDx model. The frame included an SD-IoT processor module and SD-IoT switches connected to an IoT gateway and wirelessly Connected devices. Next, a DDoS attack detection and mitigation algorithm was developed using the SD-IoT platform. To determine whether DDoS attacks have taken place in the IoT, the covariance of link layer message speed matrices at bordering SD-IoT switch ports has been utilized. Not to mention, test results demonstrated that the provided method was excellent, and the architecture had been modified to improve security in the IoT despite interacting with a range of weak devices. The similarity measure function cannot be employed for standard levels since it cannot be utilized in guided matrices.

The KEHECCS technique, which employs a signcryption approach based on position curve encryption for key lockup, was presented by Vanathy and Ramakrishnan [18]. This KEHECCS method utilizes the SSKG and GSKG strategies to promote the concept of cooperative key management. While GSKG is employed for secure group distribution, SSKG is used for private key sharing. These methods are contrasted with AES, DES, and ECC ones. The proposed method outperforms the currently used DES, AES, and ECC algorithms regarding throughput, processing costs, and network latency.

A solution to prevent wormhole assaults was put up by Jamali and Fotohi [19] using an Artificial Immune System (AIS) that can defend against various unrelated assaults while compromising the network's quality. A test package will be used and transmitted from every channel, and the target must submit a verification packet after obtaining the test package. The suggested approach consists of two steps. The packet will not reach its target, and the verification packet will not be received if the path contains wormhole hubs. Wormhole attacks typically have fewer hops in phase 2 compared to honest nodes. Therefore, the likelihood of pollutants along the way would rise when the network load was low. Which is the case when the amount of energy of the current nodes along the route increases and the overall trip duration is low.

Chen and Terzis gave research for mobile sensor nodes relying on the Log-normal path loss model in [20]. They began with the supposition that the signal intensities at close places are unrelated if they are generally distributed due to the brand perfectly. This suggests that locating the place where the packet delivery ratio is higher than a predetermined threshold can be represented as a series of Bernoulli trials. As a result, the number of searches necessary to locate a place with a high PRR is spatially dispersed. Additionally, they contended that the geometrical dispersion is preferable since it lacks lengthy tails and that, consequently, the number of searches required to identify a suitable site would typically be low, assuming that a proper proportion of locations with high PRR exist nearby. They concentrated on fluctuations in the latent space instead of fluctuations in transmission power in the temporal domain. They focused on how the log-normal model for path loss affects the deployment and movement of sensor motes.

Because of their decentralized structure and changeable architecture, Sangeetha and Kumar [21] give data packet metrics of mobile nodes in ad hoc architecture that are particularly difficult to understand. To understand the poisonous motion of the adversary, the current structure has implemented several security procedures. Designing a strategy to strengthen the interruption location framework in light of the adversary's dubious and arbitrary destructive behaviour is a statistically challenging problem. This work demonstrates a technique known as ZIDS—Zonal-based Intrusion Detection System—which removes the ambiguous methods of the malicious node by using the ability to divert the premise. ZIDS provides comprehensive protection against numerous and discrete aggressors. In contrast to active research, ZIDS performs best regarding geolocation accuracy and processing time to ensure the correct assessment of vengeful nodes and events.

The aggregate machine learning approach can identify IoT sensor dangers based on their results, according to a haven things design internet that Liu et al. [22] described. The support vector technique was chosen in an Internet of Things context with the fewest altering hyper-parameters and used to detect discriminating assaults. The system utilized a wide range of procedures, like detecting attacks, data cleaning, convnet, clustering algorithms, and internet-of-things sensing devices (IoT sensor networks) to collect internet traffic. First, the information size was reduced using the feature extraction technique, which enhanced the assault and recognition scenario. The sample size was therefore reduced using the method for feature selection. Following the prior stage with minimal hyper - parameters changing, the aggregate support vector technique was used to achieve the best results. The model was more effective, as seen by its greater accuracy of 99.40%, consistency of 99%, and F1 scoring of 99% in identifying intrusions on the IoT device's ecosystem. The method used throughout the procedure necessitated a more fabulous training time.

According to Sathiya and Gomathy [23], an attacker can obstruct access to data in MANETs by acting as an access point on the route from the origin to the destination. They present a novel method by utilizing the Beer-Quiche theoretical routing framework. In this approach, the source node monitors the accessible path at each phase, the surviving bandwidth, and the aggressor policy, collecting the data readily accessible by the preceding stage. The data packet chooses the best route for packet transmission using this data. Additionally, they recommend a suitable solenoid valve to choose from various ways from the origin to the destination. Once more, they do not consider the WANETs' DHT-based routing paradigm.

Santhi and Tamilarasi [24] proposed a technique to prevent wormhole attacks in MANET by locating the wormhole and choosing the optimum route. Using the Ad-hoc on-request Spread spectrum Shortest Path multicast routing, various paths between the origin and destination will be constructed, named "K." The data packet will then use the target's detecting package (DP) and response package to determine the time vortex path. Using the Particle Swarm Optimization (PSO) method, the data packet will choose the optimal path from the attacking player routes after identifying the time portal paths and send the data to its destination along the correct route.

According to Chaudhary et al. [26], all of the current wireless base stations must interact among themselves, self-efficacy the layout, and start the demand for info multipacks to send and receive even though portable connections are entirely bereft of any preceding structure or confirmation point. Security-wise, communication between mobile nodes via distant links renders these systems more vulnerable to internal or external assault because anybody can join and relocate the network anytime. After all, is said and done, one of the possible attacks in the portable, specially built system is a packet-lowering attack through the harmful router. In order to identify the transmission-lowering attack from the mobile organized platforms and to remove the malicious nodes to protect the resources of mobile nodes, this research focused on developing an abnormal state framework using fluffy logic.

In her article titled "A Genetic Algorithm-Based Multipath routing Strategy for Wireless Networks," Wang [27] suggests a strategy for navigating sensing devices utilizing genetic techniques that improve fault endurance and consume less energy for network elements. The numerous distance characteristics among varied different types of nodes are considered in order to obtain a practical fitness value. The range between a transmitter to a receiver and the central node is included, the number of hops from the transmitter to a receiver to the access point and the number of hops from the access point to the subsequent hop. A parametric study also supports the effectiveness of the strategy. The focus should be on privacy since it is such a crucial problem.

## 3. Proposed Methodology

This part will make a new ad hoc wireless strategy with effectiveness and safety systems. This system is flexible enough to react to actual wireless conditions. A supervising flow of packets stands on top of the semi-trusted in the Safe Routing Approach (SRA), a two-tiered approach. The proposed work briefly describes the SRA's routing process in this section, the attacker node identification method, the trust model, and the monitoring flow of packets. The Block diagram of the Suggested Safe Routing Strategy is shown in Fig. 2. (SRA). Our scheme offers confidentially, transparency, and versatility for multipath in addition to meeting the specifications of earlier methods. Additionally, it provides increased effectiveness to make the communication ecosystem more helpful.

### 3.1. Supervising the Flow of Packets

There is no active network midway structure in MANET. As a result, numerous attack nodes can readily access the channel's topology and disrupt its computing power. The source nodes and the destination network are not directly connected. To exchange the incoming packets with its relay node, the originator node sends a request for a route [39]. The packets are sent via the chosen, most efficient path from the relay node to the receiver node. An attached node may enter the route and release the attackers since the path is generated on request. An attack known as the transmission delay for diversity invaders occurs when a node waits to transmit a transmission to the following nearby node.

Typically, the development of a reliable link is required for functions like data transfer. This link is provided by the distribution method, which is based on a trustworthy link and requires a response packet within a predetermined amount of time for appropriate data packet transmission. The network latency for the diversity invader node prevents this. It displays the network latency before communication. The base station assumes that the package has not reached the destination node since the response package does not reach the source network in an instance of time, whereby Prob(A) is the package connection, Prob(f) is the network stream, and ts seems to be the time frame.

$$Prob(A) = Prob(f) * t_s \qquad (1)$$

The transmitter node continuously sends the data packet upstream, which can cause the communication network to become overloaded. As a small number of data packets are ultimately delivered, lowering the delay in packet broadcast decreases node effectiveness. This method works by having each node forward for typical broadcasting packets after a predetermined time interval and confirming that the nodes are transmitted delay for the packet by an instance of time more significant than the entrance rate across its relay node. Network characteristics like connection delay and node transmission latency occurrence determine this cutoff value. Additionally, the max amount of overburden on the path is taken into account while thinking about packet delay, where Sz(Pkt) is the length of the package, Curn is the prevailing node, S(A) is the self-association of the packet, and An is the hidden terminal detection.

$$Prob(f) = S(A) + A_n \qquad (2)$$

$$Cur_n = Cur_1 + Cur_2 + .. + Cur_n \qquad (3)$$

$$S(A) = S_z(Pkt) * Cur_n \qquad (4)$$

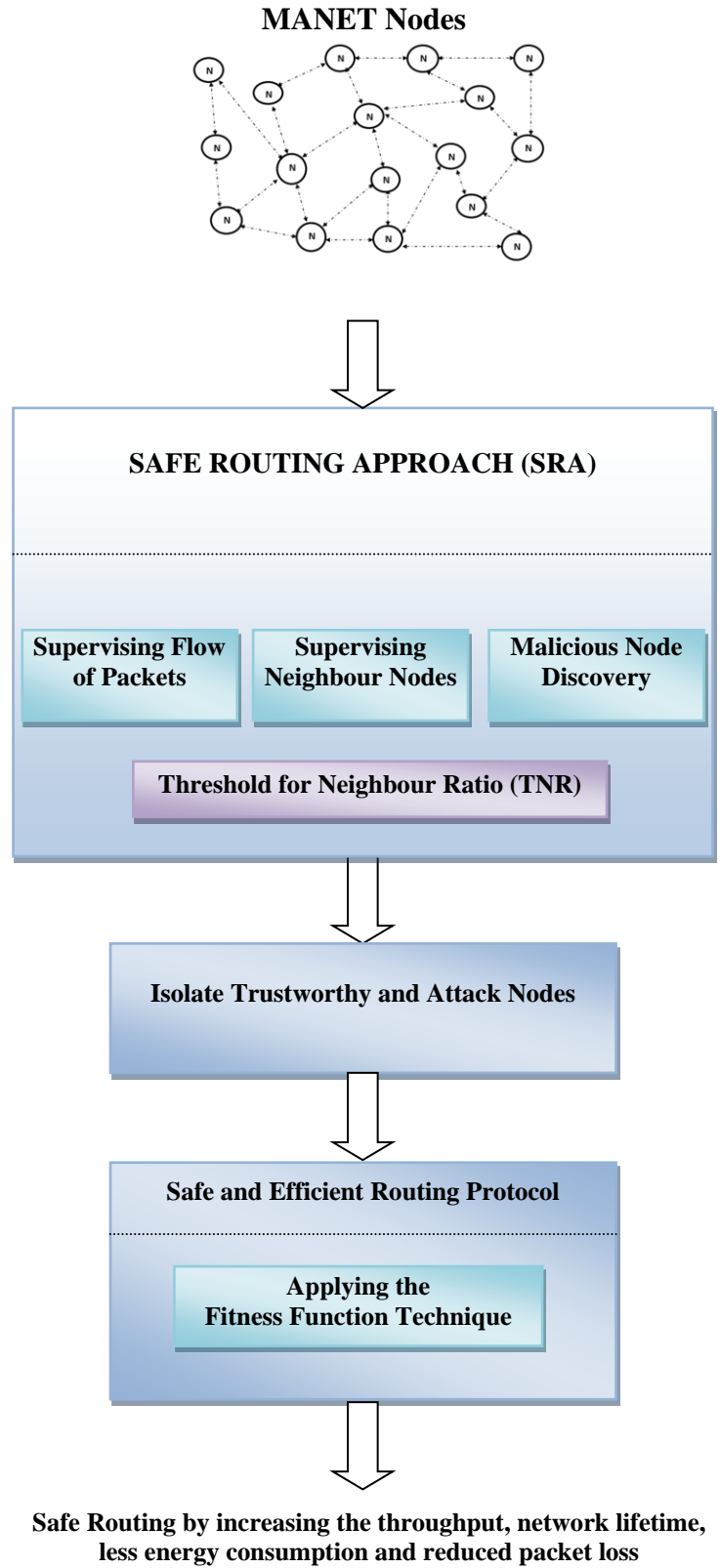

Fig. 2 Block diagram of Proposed Safe Routing Approach (SRA)

When an attack node is not accessible, the datagram transmission occurs on a route with the fewest number of nodes possible. Except when an assault node is discovered on the shows the locations, the node next to the assault node does not take the intended route; instead, it chooses constructive trust, regardless of availability. As a result, the attacker node will never enter the transmitting path since it loses the confirmation signal from the assault node [40]. The node inside the mobile network is forced to choose a higher sequence number route rather than the routing process, which is the shortest range route with the assault node. Performance data shows that the transfer rate is higher using the route without an attacker node, and the point-to-point transmission delay is reduced. Consequently, the net display improves.

### 3.2. Malicious Node Discovery

The method for determining whether each network device is an assault site or not is among the most energy-intensive ways, adding latency for the nodes in the network. Attack nodes raise the number of neighbours, which improves connection and leads to erroneous RTT. Consequently, a straightforward method known as the Threshold for Neighbour Ratio has been utilized (TNR). To find the assault networks in MANET, it will check a node's neighbour value with all of its neighbours. The nodes will be aware of their neighbours using neighbour identification operations. The nodes then determine the neighbour ratio, which is the proportion of its neighbour number to the mean neighbour amount of all its neighbours.

As shown below in algorithm 1, in which the complete network NW contains nodes N and their neighbours set Ns, the neighbour ratio (Nr) will then be matched with the Minimum for Neighbour Ratio to evaluate if assault node identification is necessary or not. Then, to carry out an out-of-band detection mechanism, nodes with neighbourhood ratios greater than the Threshold for Neighbour Ratio would be included in the suspicious list [41]. The cutoff for Neighbour Ratio's primary objectives is to decrease latency and energy costs while executing assault detection techniques.

Algorithm 1 - Threshold for Neighbour Ratio (TNR)
Input - Set of nodes
Output - TNR (Threshold for Neighbour Ratio)
for *each node Nj in N*
  do
    $sj = |Sj|$ (neighbour amount of *Nj*);
    For *each node Nj, ∈ Sj* do
    $sk = |Sk|$ (neighbour amount of *Nk*);
    x= 0;
    x= x+ *sk*;
    To determine the average quantity of *Nj* neighbours
    Then $sj = x/sj$
    To discover *Nj* neighbour ratio $TNRj = sj/sj$
 if   $TNRj > TNR$ then

    set *Nj* to alleged nodes - *A*
  end

The system's penetration is assessed when the optimal heads are determined. The next step is to examine these nodes' hopeful components from the state's and the target clients' perspectives. The suspect's identification has been identified, and the infiltrating node is not permitted to connect to the network. To assess the existence of invaders, the sink node's specified limit is employed. The threshold quantity may be chosen at the supervisor's choice according to the system's needs. The value is typically 0.5J in most instances.

The remaining power spectrum of a specific node is used to calculate a route's transmitting capacity and connectivity condition. A node loses access to a system when its power level is low because it is less likely to find another unit nearby. However, a node's transmit power might be improved if its energy is higher. The fundamental objective of vulnerability scanning is to keep a secure communication link while consuming minimal energy and sending data as quickly as possible.

Nm indicates how many of the network's nodes are acting improperly. The likelihood of malfunctioning the network's nodes, therefore, is provided by:

$$Prob_m = \frac{N_m}{N_n} \quad (5)$$

where Nn is the network's total amount of nodes. When none of the nodes that make up a route is "misbehaving," the path is said to be safe. Let h represent the typical quantity of hops on a route. It is possible to calculate the likelihood of finding a safe path from the origin Sr and target Dt as follows:

$$Prob_{sec}(S_r, D_t) = \left(1 - \frac{N_m}{N_n}\right)^{h-1} \quad (6)$$

The value of h should be evaluated to understand how the quantity of malfunctioning nodes affects the likelihood of discovering a safe route. If h is the mean number of hops in a path and d(nda, ndb) is the average length between two nodes within the network. Following that, the following average distance is reported among source Sr and destination Dt:

$$Dist.(S_r, D_t) = h * Dist(n_{da}, n_{db}) \quad (7)$$

Allow a node's communication range to reach between points (ao, bo) and (a, b). The node's and range's transmitting circle diameter can be calculated as follows:

$$T_r = \frac{\sqrt{(a-a_0)^2 + (b-b_0)^2}}{2} \quad (8)$$

$$N_a(T_r) = \frac{N_n}{(a-a_0)(b-b_0)} \int_0^{T_r} 2\pi dr \quad (9)$$

Smart adversaries may pose as innocuous nodes by transmitting traffic before initiating packet-smashing assaults to increase the trust level in nearby nodes. Nevertheless, during the route identification process, packet loss attackers perform sequence number attacks that may produce a particular type of pattern in generating some parameter value in the datagrams.

SRA uses a trend-finding approach to examine the captured parameter value from the eavesdropped control messages. Two sliding windows are used to store the parameter value: The target node's identification, present moment, node density, and sequence number are all recorded in the initial sliding window (SL1), and the target node's individuality, current time, and the differential in between sequence number figures of the obtained piece and that of the correlating request message are all recorded in the main sliding window (SL2).

The collected data is analyzed by an algorithm using the method of the Frequent Variations model, which then determines whether the neighbour node has any known attack patterns. This approach strengthens the route-finding process by segregating the prohibited enemies who might subsequently execute packet loss assaults [42-44]. It needs to be emphasized. Nevertheless, that attacker node might keep dropping packets until the monitoring node has filled all of its sliding windows and determined that they belong on the blocklist.

### 3.3. Safe and Efficient Routing Protocol

The proposed method aims to identify the paths with the highest probability of success. The requirement for quick reaction programming will also indicate the optimizing algorithm's result, and the chosen paths for the computation transit in MANETs constitute the solution [45]. The energy remaining in the nodes, the network's capacity, and the path's availability affect how healthy away it is. Consequently, the fitness function is an optimizing variable. Using the Fitness Function approach to optimize energy consumption, the suggested study is highly particular on the issue of energy usage in MANET.

In multi-hop networks, the fitness function determines the best route from the origin to the destination to conserve energy. The fitness function is a method of optimization. According to the research's objectives, the fitness function identifies the most crucial element to the optimization process, which could be any number of factors [46]. The fitness factors in MANET are typically bandwidth, energy, range, and latency. The fitness function employed in this study is a component of the SRA algorithm. In the suggested task fitness function, the alternate route is optimized in case the main path fails or an attack node is detected. The fitness function is represented mathematically by the following equation.

$$Fit = \frac{1}{3}(E_g + t_p + P_c) \quad (10)$$

$$= \frac{1}{\sum_{a=1}^{len_i - 1} lc_{bi}(a) + bi(i+1)} \quad (11)$$

The nodes of the route are used to calculate Eg, Tp, and Pc, representing energy, capacity, and path connectivity, respectively. leni stands for length, and lcbi stands for link cost.

## 4. Performance Analysis

Utilizing MATLAB R2021b, the performance of the Safe Routing Approach (SRA) is evaluated. In terms of Packet Delivery Ratio (PDR), network performance, energy usage, and recognition of attack nodes, the scheme is contrasted with AIS: Artificial Immune System [19], ZIDS: Zone-based Intrusion Detection System [21], and Improved AODV [25]. By raising the fraction of malicious nodes, the efficiency of a net is also evaluated. The experiment's simulation settings are displayed in Tab. 2. Intel Pentium Core i5 with a clock speed of 2.8 GHz, 16 GB of RAM, and 64-bit Windows 8.1 have been employed. The same platform mimics each compared procedure to ensure constant results collecting. To evaluate the effectiveness of the proposed methods, a random sample of origin nodes of size varying from 25 to 150 nodes was repeated to use the simulation system depicted in Table 1. All source nodes were configured to transmit constant bit rate (CBR) data packets at arbitrary speeds ranging from 25 m/s to 30 m/s. The random waypoint (RWP) technique produces various nodes. It takes only 700 s of simulation time to assess a network's complexities, delay, and traffic.

**Table 1. Simulation Parameters**

| Parameters | Values |
|---|---|
| Number of Nodes | 25 - 150 |
| Simulator | NS 2.3 |
| Initial Energy | 100 J |
| Simulation Time | 700s |
| Bandwidth | 2 Mbps |
| Packet size | 512 bytes |
| Speed of the node | [25-30]m/s |
| Mobility Model | Waypoint |
| Wireless Range of Transmission | 250s |
| Protocol used for routing | RIFA |
| Traffic category | CBR |
| Area of Simulation | 1200m |
| Pause Time (s) | 10s |
| Node Locations | Random |

### 4.1. Packet Delivery Rate

It might be characterised as the proportion of received packets to all packets sent. It contains packets for RREP and RREQ. The PDR of the connection ought to be high. A network with a high PDR is more trustworthy.

Table 2. PDR with changing % of Attack Nodes

| % of Attack Nodes | AIS [19] | ZIDS [21] | Improved AODV [25] | Proposed Approach |
|---|---|---|---|---|
| 10 | 0.88 | 0.89 | 0.9 | 0.96 |
| 20 | 0.88 | 0.83 | 0.87 | 0.93 |
| 30 | 0.74 | 0.79 | 0.8 | 0.84 |
| 40 | 0.67 | 0.74 | 0.71 | 0.77 |
| 50 | 0.54 | 0.58 | 0.6 | 0.67 |
| 60 | 0.45 | 0.47 | 0.51 | 0.58 |

### 4.2. Network Throughput

Network throughput is the proportion of received packets at a target. Throughput and PDR might be mixed up. Consider a scenario where packets are appended to the cache during the early phase owing to wait time. PDR will not be impacted because the packages have not yet arrived in the system, but bandwidth will be decreased. However, if packet delivery fails twice or more, PDR will drop, and there will not be much of an impact on capacity.

Fig. 4 compares network bandwidth with AIS, ZIDS, and Enhanced AODV. The Safe Routing Approach (SRA) offers more efficiency than current methods. SRA is superior because it selects an efficient routing strategy while maintaining a route's trust level, energy, and hop count. It effectively gets rid of broken nodes. The SRA trust factor is additionally dynamic. CH is chosen based on high-quality links. SRA has higher throughput than currently used methods as a result.

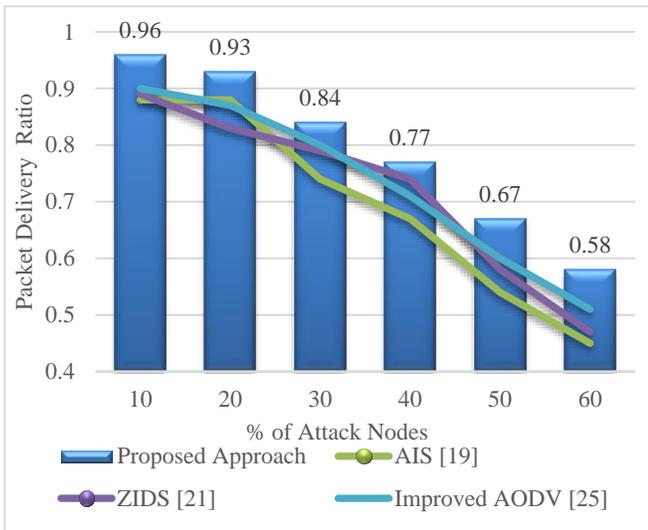

**Fig. 3 PDR with changing % of Attack Nodes**

Fig. 3 displays a comparison of the proposed strategy with existing methods. Because hybrid routing is used in the proposed strategy, it can be shown that the packet delivery rate is superior to that of existing methods. The cluster-level maintenance of routing tables increases the packet delivery ratio. Additionally, the shortest path is chosen while choosing the route discovery in the suggested method utilising hop count, which significantly decreases wait time and hence boosts the packet delivery ratio. This standard represents the level of the proposed method from an origin to a destination.

The proposed method's effectiveness increases with the data packet delivery pace. Let PDR represent the data packet delivery performance, which is determined using the following formula:

$$\text{Packet Delivery Rate} = \frac{N_{pr}}{N_{ps}} * 100\% \qquad (12)$$

Where $N_{pr}$ is the number of packets received while $N_{ps}$ is the number of packets transmitted.

Table 3. Throughput with changing % of Attack Nodes

| % of Attack Nodes | AIS [19] | ZIDS [21] | Improved AODV [25] | Proposed Approach |
|---|---|---|---|---|
| 10 | 0.87 | 0.92 | 0.91 | 0.96 |
| 20 | 0.88 | 0.86 | 0.87 | 0.91 |
| 30 | 0.74 | 0.77 | 0.74 | 0.79 |
| 40 | 0.67 | 0.66 | 0.66 | 0.67 |
| 50 | 0.51 | 0.58 | 0.55 | 0.57 |
| 60 | 0.43 | 0.47 | 0.47 | 0.51 |

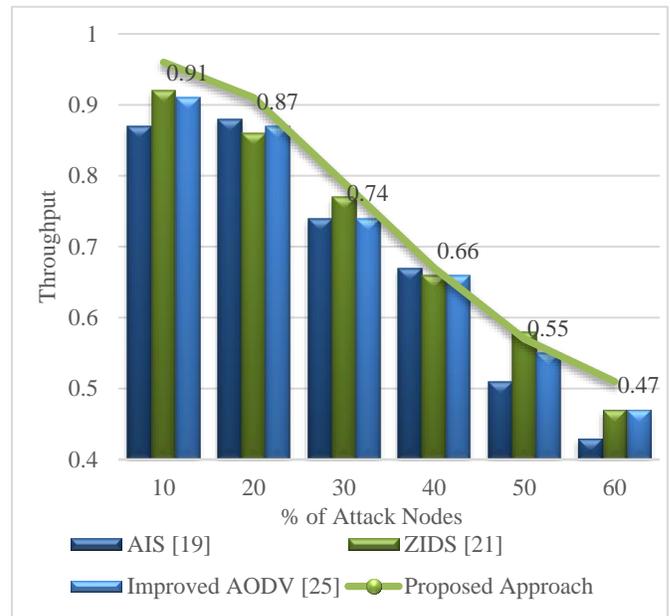

**Fig. 4 Throughput with changing % of Attack Nodes**

Table 4. Energy Consumption - Simulation Time

| File Size (Kb) | AIS [19] | ZIDS [21] | Improved AODV [25] | Proposed Approach |
|---|---|---|---|---|
| 10 | 38 | 31 | 37 | 31 |
| 20 | 66 | 65 | 69 | 59 |
| 30 | 65 | 68 | 72 | 61 |
| 40 | 75 | 90 | 84 | 75 |
| 50 | 87 | 96 | 102 | 80 |
| 60 | 99 | 104 | 124 | 93 |

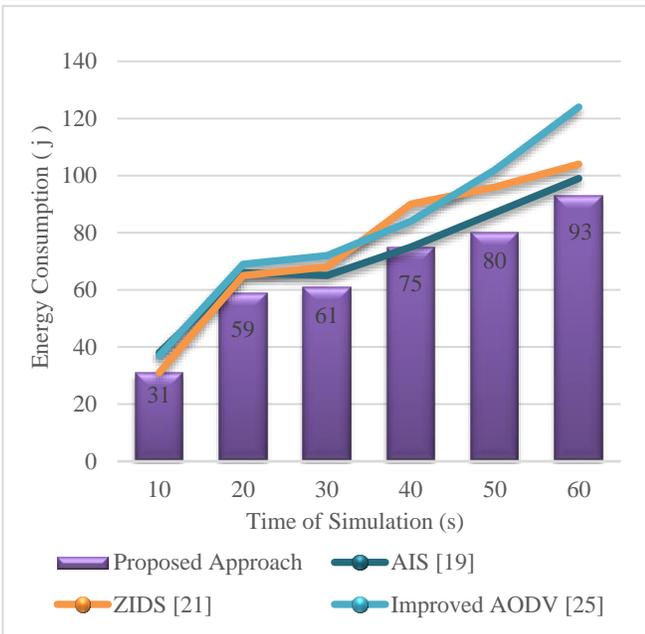

**Fig. 5 Energy Consumption - Simulation Time**

Table 5. Detecting Attack Nodes with changing % of Attack Nodes

| Detecting Attack Nodes (%) | AIS [19] | ZIDS [21] | Improved AODV] [25] | Proposed Approach |
|---|---|---|---|---|
| 10 | 0.85 | 0.87 | 0.9 | 0.94 |
| 20 | 0.8 | 0.83 | 0.85 | 0.9 |
| 30 | 0.7 | 0.75 | 0.78 | 0.84 |
| 40 | 0.67 | 0.71 | 0.7 | 0.74 |
| 50 | 0.59 | 0.58 | 0.63 | 0.68 |
| 60 | 0.46 | 0.49 | 0.51 | 0.54 |

### 4.3. Energy Consumption

The total energy network nodes used during the scenario are referred to as energy consumption. This is achieved by determining each node's energy level after the experiment and considering its residual energy [36-38]. Econs will stand in for energy usage, which can be computed as

$$Econs = \sum_{a=1}^{B}(Ei(a) - Er(a)) \qquad (13)$$

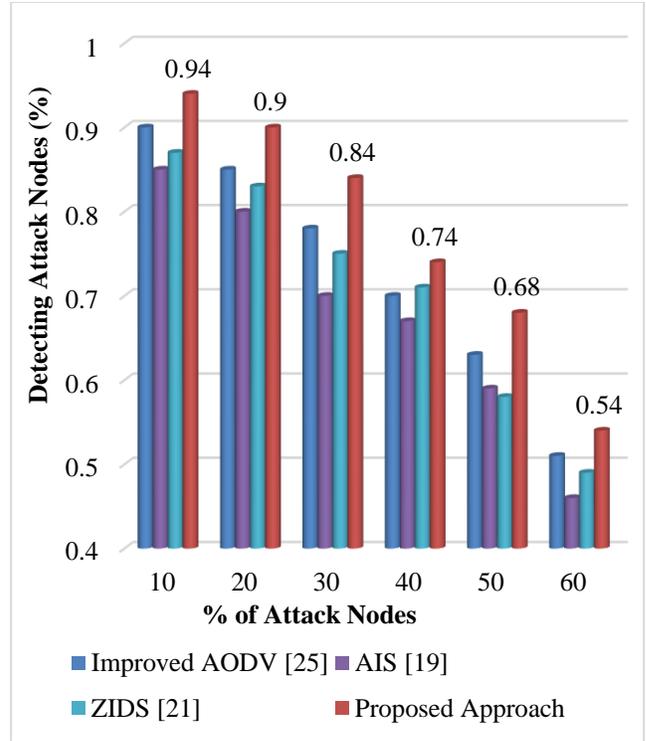

**Fig. 6 Detecting Attack Nodes with changing % of Attack Nodes**

It is the proportion of the total amount of energy used by all nodes to all nodes. Figure 6 displays the typical energy used by various numbers of nodes for proposed and existing designs. If we assume that a packet requires a fixed amount of energy to be transmitted, the frequency of communications directly affects the network's average energy utilisation. The suggested system chooses the shortest path, which lowers the number of transmissions and overall energy usage. Additionally, using routing information in clusters will save energy that is used to discover routes repeatedly. The suggested approaches are AIS: Artificial Immune System [19], ZIDS: Zone-based Intrusion Detection System [21], and Improved AODV [25], and their energy consumption is displayed in Fig. 6. In contrast to existing protocols, which employ and over 35 joules for 10 seconds and 95 joules for 60 seconds, the proposed technique utilizes 31 joules for 10 seconds and 93 joules for 60 seconds. The proposed technique consumes less energy than several alternative methods.

### 4.4. Identification of Attack Nodes

Attack nodes are seeded into the system at varying percentages to verify their identification. Several assault nodes are visible in Fig. 7, with variable percentages of malicious nodes.

The low influence on trust value is due to the trust management function. Because direct trust is based on three elements and SRA uses two degrees of criterion, this is the case. In addition to this, SRA offers a flexible time. Fig. 7 demonstrates that SRA outperforms existing approaches concerning PDR, energy consumption, network speed, and attacker node detection. Because the SRA is dynamic and senses trust in time slots, it can identify nodes that initially displayed appropriate behaviour to avoid detection. The trust value, lowest hop count, and remaining energy are also considered when choosing a route to transport a packet. As a result, an effective short way that is also durable is chosen.

## 5. Conclusion

The most common threats and attacks, such as wormhole attacks, can affect MANETs. The efficiency of the wireless network is undermined, and most routing protocols are disrupted by wormhole attacks, which are challenging issues that track packets through one spot on the system and route it to another. To identify attackers using various attack patterns when they start packet-dropping attacks, we developed a novel trust-based method called SRA for MANETs. In order to improve the effectiveness of activities, the plan aimed to segregate the enemies early on. A suggested scheme's significance is compared to existing ones like AIS, ZIDS, and Improved AODV regarding Packet Delivery Ratio (PDR), network throughput, energy usage, and attack node identification. The findings demonstrate that the suggested design enhanced throughput, average energy use, and PDR.

Additionally, SRA has a minimal effect on malicious nodes' trust value. It is, therefore, ideal for a wide range of applications. Future SRA modifications will enable it to be utilized for IoT-based apps.